\documentclass[sigconf, nonacm]{acmart}
\AtBeginDocument{%
  }

\setcopyright{none}

\usepackage{multirow}
\usepackage{listings}

\newcommand{\embedding}{DashCLIP}

\begin{document}

\title{\embedding{}: Leveraging multimodal models for generating semantic embeddings for DoorDash}

\author{Omkar Gurjar, Kin Sum Liu, Praveen Kolli, Utsaw Kumar, Mandar Rahurkar}

\email{{omkar.gurjar, kinsum.liu, praveen.kolli, utsaw.kumar, mandar.rahurkar}@doordash.com}
\affiliation{
  \institution{DoorDash, Inc.}
  \country{United States}
}

\renewcommand{\shortauthors}{Gurjar et al.}

\begin{abstract}
Despite the success of vision-language models in various generative tasks, obtaining high-quality semantic representations for products and user intents is still challenging due to the inability of off-the-shelf models to capture nuanced relationships between the entities. In this paper, we introduce a joint training framework for product and user queries by aligning uni-modal and multi-modal encoders through contrastive learning on image-text data. Our novel approach trains a query encoder with an LLM-curated relevance dataset, eliminating the reliance on engagement history. These embeddings demonstrate strong generalization capabilities and improve performance across applications, including product categorization and relevance prediction. For personalized ads recommendation, a significant uplift in the click-through rate and conversion rate after the deployment further confirms the impact on key business metrics. We believe that the flexibility of our framework makes it a promising solution toward enriching the user experience across the e-commerce landscape.



\end{abstract}

\maketitle


\begin{figure*}[h]
  \centering
  \includegraphics[scale=0.55]{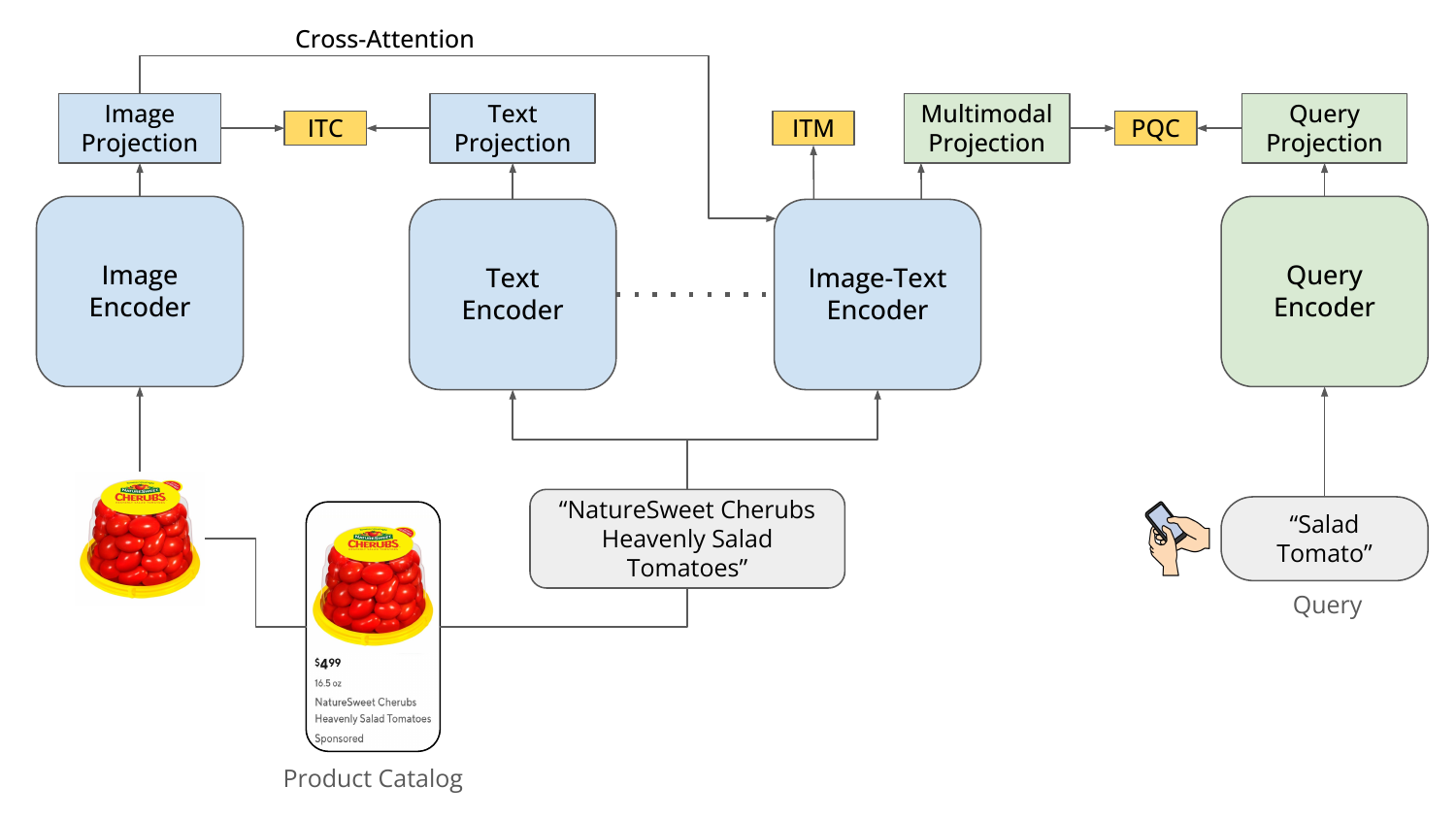}
  \vspace{-1.5mm}
  \caption{Model Architecture and Training Objectives of \embedding{}. We perform the training in two stages. Stage 1 (colored blue): image and text uni-modal encoders are trained using the Image-Text contrastive (ITC) loss, and multi-modal image-text encoder is trained using the Image-Text matching (ITM) loss. Stage 2 (colored green): We train the multi-modal projection layers and the query encoder using the Product-Query contrastive (PQC) loss. Dotted line represents shared weights.}
  \label{fig:model_arch}
  \vspace{-2.5mm}
\end{figure*}

\section{Introduction}

Personalized recommendation \cite{cheng2010personalized, pancha2022pinnerformer, zhang2024scaling, baltescu2022itemsage}, ubiquitous in e-commerce and social network platforms on the internet, is to find out relevant entities given the user preferences and context-dependent needs. This aligns with the business objective of creating user value and fostering long-term customer loyalty. Deep learning models have emerged as the foundation of modern machine learning systems in the space, excelling at capturing complex relationships between user preferences, product attributes, and contextual signals \cite{cheng2016wide, guo2017deepfm, wang2017deep, wang2021dcn}. A key strength of deep learning models is their ability to learn embedding representations for categorical features while also leveraging pre-trained embeddings generated by more sophisticated models that are infeasible to serve online at web scale. In practice, deploying embedding-based approaches across various e-commerce applications presents several challenges: (1) Embeddings generated from historical user interactions may fail to capture the rich semantic structure of product attributes, resulting in biased exposure and diminished diversity. (2) Pretrained embeddings may struggle to generalize effectively to out-of-distribution scenarios, restricting their applicability across diverse product and user tasks.


In this work, we present an embedding generation and alignment framework that we develop within the context of \textbf{DoorDash}, a large-scale online e-commerce platform which encompasses the delivery of groceries, retail products, alcohol, electronics, pharmaceuticals, and more. Our framework, named as \textit{\embedding{}}, is built on top of the recent methodology of pre-training multi-modal models using contrastive learning \cite{radford2021learning, li2022blip}.  \embedding{} leverages these key design choices to create semantic entity representations which are generalizable to different functional requirements:


\noindent \textbf{Multi-Modality Encodings}: In e-commerce, products are typically associated with both textual and visual information. Instead of processing these modalities separately, it is essential to integrate the rich, yet unstructured, data from both sources to create a unified representation. In this work, we leverage contrastive learning on the product catalog of DoorDash to approximate human-like understanding of products, capturing the complementary information from each modality.

\noindent \textbf{Domain Adaptation}: Transfer learning provides a solution by enabling fine-tuning of the off-the-shelf models for domain-specific applications, as demonstrated in prior work on fine-tuned image embeddings \cite{zhai2019visual} and fine-tuned text embeddings \cite{zhuang2019pintext2}. To achieve this, we perform continual pre-training on a multi-modal transformer-based model using contrastive learning on our internal catalog data, leveraging product images and titles as inputs to adapt to the platform's specific catalog distribution.

\noindent \textbf{Embedding Alignment}: To ensure seamless integration with downstream applications, the architecture must be adaptable to accommodate new task objectives. For search recommendation, we introduce a second stage of alignment for a dedicated query encoder. It is designed to generate query embeddings that are learned in the same space as product embeddings, enabling more effective retrieval and ranking.

\noindent \textbf{Relevance Dataset Curation}: For a product search application, product-query relevance is crucial to match the user intents and product offerings. Most prior systems resolve this connection through historical user engagements like clicks or conversions, which are prone to position and selection bias. To address this, we curate a high-quality relevance dataset that combines internal human annotations with knowledge from large language models, providing robust supervision for embedding alignment.

The remainder of this paper is structured as follows: Section 2 covers the data collection and curation; Section 3 details the architecture of multi-modal encoders, along with the training procedure; Section 4 offers a thorough evaluation in search recommendation; Section 5 discusses potential e-commerce applications; and Section 6 \& 7 concludes and reviews related work.

\section{Data Collection}
\label{datasets}

\subsection{Catalog Dataset}
\label{catalog-dataset}
The product catalog contains various types of information related to products available on DoorDash. This includes merchant-provided and internally collected data about the product. We curate a list of around 400k products and use their catalog data for our continual pre-training and evaluation tasks. While the catalog contains different fields, we find the following to be important due to their semantic value: \textit{1) Title:} The product name as shown to the user. \textit{2) Image:} A standardized product image. \textit{3) Description:} Usually, a single sentence elaborating some additional information, e.g. dimensions for packaged goods. \textit{4) Detail:} Usually, a few lines or a paragraph explaining the product features which are more verbose compared to the description. \textit{5) Aisle Category:} The product category classified with internal taxonomy, e.g. drinks, snacks.  

For this paper, we only utilize the product title and image to train the embeddings. We make this decision for two reasons. First, the title and image are likely to receive the largest user attention on the platform as only these two fields are visible to the users in the category or search results page. 
Second, while the description and detail sometimes contain additional information, we observed some data quality issues with these two fields. The detail is often framed like an advertisement for the product which makes it quite noisy to use without further processing. Additionally, from an exploratory data analysis conducted on selected 13k products, we found that the description and detail to be missing for 32\% and 20\% of the products respectively, which raised data coverage concerns. We also use aisle category as part of the sampling strategy for contrastive learning and the additional downstream task described in Section \ref{prelim-eval-category}. 

\subsection{Query Product Relevance Dataset}
\label{query-prod-dataset}

\lstset{
  frame=single,
  breaklines=true,
  columns=fullflexible,
  xleftmargin=0pt,
  numbers=none,
  showstringspaces=false
}

In order to align the query embedding and product embedding in the same space, we require a relevance dataset that assigns a relevance label from \{0: irrelevant, 1: moderately relevant, 2: highly relevant\} to each <query, product> pair. We refer to these labels as \textit{IR}, \textit{MR}, and \textit{HR}. We used a hybrid approach to utilize both human annotation and a LLM to create such a dataset. Specifically, we have 700k human-annotated relevance labels, which is used to fine-tune a language model (GPT 3.5 as the final choice) to grade the relevance of any new <query, product> pair. In total, there are 32 million LLM-generated relevance labels for such pairs which cover 
more than 99\% of our search volume. The distribution of the labels is \{IR: 69\%, MR: 20.5\%, HR: 9.9\%\}. This process was done in late 2023 so GPT3, 3.5, 4 were evaluated. For our specific relevance task, we found that more sophisticated models may not guarantee enhanced performance, but fine-tuning has a notable improvement. Thus, we fine-tuned with 600k human annotations and evaluated it with the remaining 100k to select the best version for inference. 
The fine-tuning prompt specified the role of the language model and the definition of different relevance labels. After the initial prompt, a list of <query, product> examples with labels are also appended.

\section{\embedding}

\subsection{Model Architecture}
The goal of \embedding{} is to create generalizable product representations that can be applied across various downstream applications. 
To achieve this, we adopt the vision-language pre-training framework outlined in \cite{li2022blip}, which consists of two uni-modal encoders (one for image and one for text) and a third image-grounded text encoder. These encoders work together to generate product representations. To address scenarios involving user search, such as the search recommendation experiments discussed in Section \ref{ctr-experiment}, we incorporate an additional query encoder. This query encoder utilizes a text-only transformer to process normalized free-text search terms provided by the user. The final model architecture is illustrated in Figure \ref{fig:model_arch}.

\subsection{Model Training}
We initialize the image-text Product Encoders from a pre-trained checkpoint BLIP-14M \cite{li2022blip}. The authors also make available the checkpoints which are fine-tuned on COCO \cite{lin2014microsoft} and Flickr30K \cite{flickr30k} datasets, however, we don’t opt for these as the fine-tuning datasets comprise of open-world and human images which are quite different from product images. After initialization, we adopt a two-staged training methodology: 

\subsubsection{Continual Pre-training of Product Encoders}
In this first stage, we continue the pre-training of the Product Encoders on the 400k raw product images and titles from our Catalog Dataset mentioned in Section \ref{catalog-dataset}. By exposing the encoders to the product data from the platform's catalog, the encoders will adapt to the characteristics and patterns of the product domain. This optimization stage aims to minimize the contrastive loss between image and text embeddings (ITC), and the matching loss of the image-text representations (ITM). We follow the same soft label creation and negative sampling strategies from \cite{li2022blip, li2021align}. In order to avoid overfitting the encoders, we freeze the first 8 layers of image and text encoders and early-stop at the lowest evaluation score of the matching loss.

\subsubsection{Aligning Query and Product Encoders}
In the second stage, we initialize another Query Encoder from the text encoder of BLIP. Then, we align the query embedding with the product embedding by minimizing a contrastive loss in the projection space of the image-text Product Encoder and text-only Query Encoder. Again, we freeze the first 8 layers of the Query Encoder. Inspired by the SimCSE \cite{gao2021simcse} contrastive loss, we design the Product-Query contrastive loss (PQC) as
\begin{equation}
  \sum_{i=1}^{B} - \log \left( \frac{e^{sim(C_i, Q_i^{+})/\tau}}{e^{sim(C_i, Q_i^{+})/\tau} + \sum_{j=1}^{N} e^{sim(C_i, Q_{ij}^{-})/\tau}} \right)
\end{equation}
where $C_i$ is the multi-modal hidden representation of the $i$-th product, $Q_{i}^{+}$ is the positive (relevant) query for the $i$-th product, $Q_{ij}^{-}$ is the $j$-th negative query among the $N$ negative samples for the $i$-th product. We average this loss over the batch size $B$. $sim$ is the cosine similarity function, and $\tau$ is the temperature parameter.

\begin{figure}[h]
  \centering
  \includegraphics[width=0.8\linewidth]{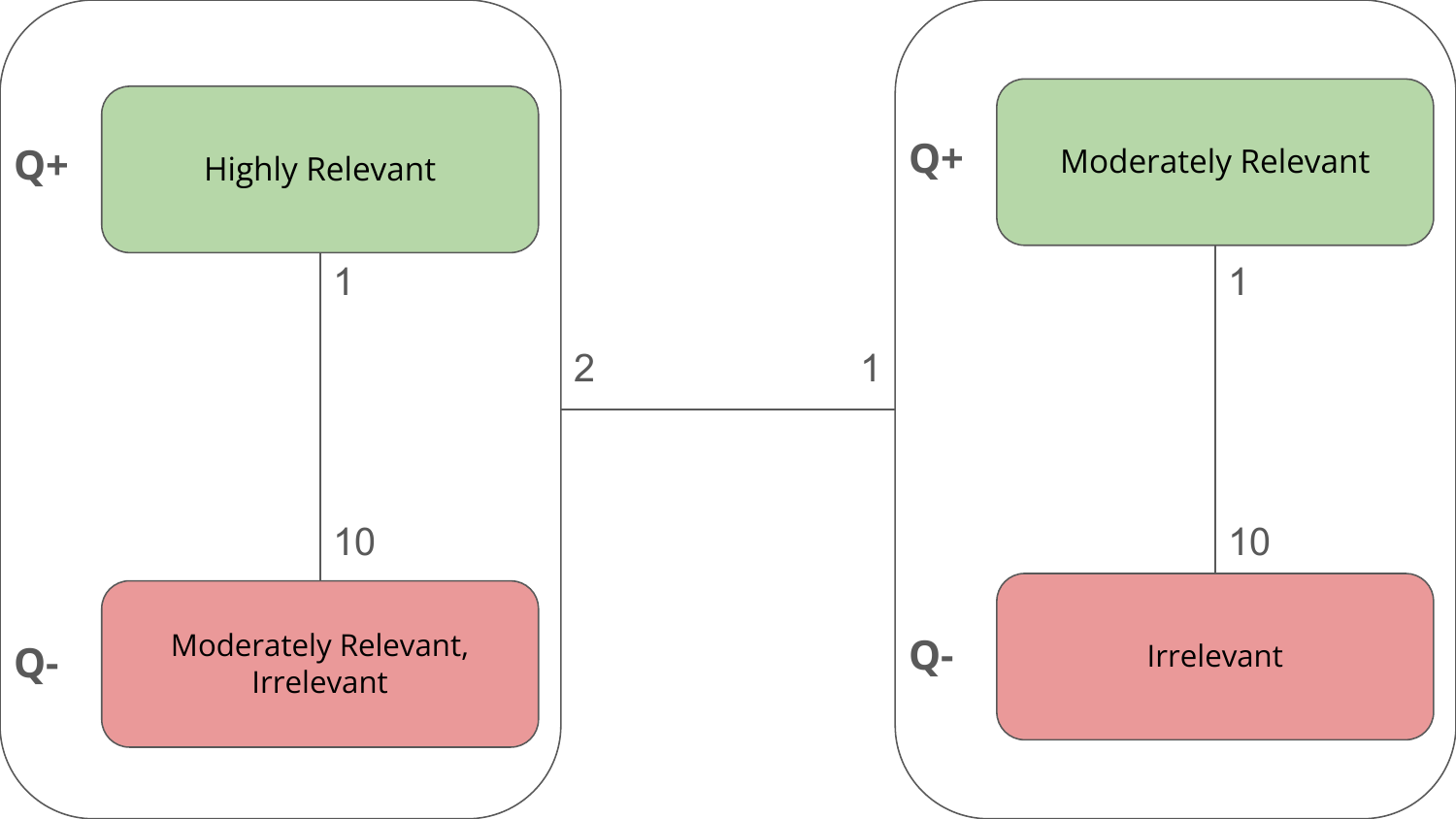}
  \caption{Sampling strategy from Query Product Relevance dataset for training the PQC objective. The numbers represent the relative frequency ratio between respective relevance types of queries. }
  \label{fig:product_query_dataset}
\end{figure}

We notice that the in-batch negative strategy used in \cite{gao2021simcse} is not compatible with our setting since we can't guarantee that all other queries in a batch would be irrelevant to a given product. Further, we also need to map our three relevance levels to the binary positive/negative labels here. To facilitate the Product-Query contrastive loss, we create tuples like (product, positive query, [negative queries]) from the Query Product Relevance Dataset described in Section \ref{query-prod-dataset}. The sampling procedure is as follows: 

First, we retrieve products with at least one moderately or highly relevant query. For each of these 24k products, we randomly sample at most 110 positive examples among the HR or MR queries in the ratio of 2:1 to focus on the more informative but infrequent highly relevant queries. On average, each unique product will be paired with 50 positive queries. For each such (product, positive query), the next step is to sample a list of 10 negative examples. If the positive query is HR, we look for negative queries from the set of IR or MR queries. If the positive query is MR, we only sample negatives from IR queries. In cases where we cannot find enough negatives from the relevance dataset, we generate proxy negatives by using highly relevant queries for another random product from a different aisle category (a HR query for a product from "Snack" can be considered irrelevant for "Medicine" products). Also, when all 10 sampled negatives are from IR queries, we randomly replace one with this proxy negative to promote more diversity. This generates a total of 1.4 million entries of (product, positive query, [negative queries]). We find that this sampling strategy strikes a good balance between hard negatives (HR vs MR queries) and diversified examples (IR and random queries) for the contrastive loss. A visual representation of the dataset preparation is depicted in Figure \ref{fig:product_query_dataset}.

\subsubsection{Training cost}
\label{cost}
We perform these two training stages to obtain the final encoders as the inference model to generate \embedding{} embeddings for any product or query. The entire training was run on 16 Nvidia Tesla T4 (16 GB) GPUs. The total training time for one experiment run was roughly 10 hours. We are using AWS EC2 instances and the estimated cost for a single training run was $\approx$150 USD. To obtain the relevance labels, inference was done on OpenAI's GPT 3.5, with an estimated cost of around 42k USD. Note that these labels are also being used for other applications within DoorDash's ads funnel.

\section{Search Ads Recommendation Experiments}
\label{ctr-experiment}

The \embedding{} product and query embeddings are generalized for different downstream applications. In this section, we focus on the task of ads recommendation, which aims to find the most relevant advertisement for a user engaging the search surface of the platform. We are going to investigate how \embedding{} performs in the retrieval and ranking stages of the ads recommendation funnel, with comparisons with various common embedding approaches and ablation studies to investigate the best model architecture to make use of such embeddings. At the end, we deployed the best ranking candidate in the production environment to serve real users and verify the business impact. Highlights of the evaluations are as follow:


\begin{enumerate}
  \item Off-the-shell vision-and-language models are not able to retrieve relevant products precisely. Alignment between the product and query in \embedding{} encoders is necessary to understand the e-commerce context. 
  \item \embedding{} embeddings outperform those learned through a randomly initialized embedding lookup matrix during end-to-end model training in ranking experiments. This highlights that utilizing the semantic information of products and queries to pre-train the embeddings provides a strong prior for downstream applications.
  \item The derived sequence features, capturing users' engagement signals using product embedding contributes to further improvement to the ranking model, which shows the flexibility of building advanced features using \embedding{}. 
\end{enumerate}

\subsection{Retrieval}

In the retrieval task, we leverage the embedding of a user's query to perform a K-nearest neighbor search in the embedding space of the product to create a ranked list of potential relevant candidates for the next downstream selection (ranking etc). Both effective representations and alignment between them are essential to retrieve relevant items at top positions. For subsequent evaluation, we used the hold-out subset of Query Product Relevance Dataset, which consists of <query, product> pairs that are not used in the training of \embedding{}, as the seed queries and corresponding retrieval pool of product candidates.

Methods to compare: For CLIP \cite{radford2021learning}, we used the ViT-B/32 Transformer architecture as the image encoder for product image and the masked self-attention Transformer as the text encoder for both user query and product title. The encoders are initialized from Hugging Face's checkpoint. For BLIP, we used the unimodal image and text encoders in the same way after the final projection layers, thus representations are aligned in the same space by the contrastive loss. Note that in BLIP, the representation from the image-grounded text encoder is not aligned with the unimodal encoders. We used the pretrained BLIP-14M checkpoint \cite{li2022blip}. For FLAVA \cite{singh2022flava}, the setup is similar to BLIP and we used the checkpoint from Hugging Face. For \embedding{}, the natural choice is the embedding from the image-text encoder and query encoder which is aligned by Product-Query contrastive (PQC) loss. 

From Table \ref{table:retrieval}, \embedding{} outperformed all baselines by significant gains, demonstrating the effectiveness of product-query alignment in our proposed framework. General LLM, while much better than random with the power of world knowledge, lacks the specificity of the e-commerce domain and often fails for short but specific queries. Surprisingly, CLIP (Image) performs notably well in terms of precision. We found that for many branded queries (such as `celsius drink'), the image encoder is able to recognize the text on the product image to perform keyword matching as a zero-shot OCR task. We also experimented with multi-modal embeddings of BLIP and FLAVA. Since there is no alignment between this representation and the text-only encoding of the query, the performance is significantly worse by default.

\begin{table}[h]
\resizebox{\linewidth}{!}{
  \begin{tabular}{lccc}
    \toprule
    \textbf{Embedding Variant} & \textbf{Graded Rel} & \textbf{Recall@K} & \textbf{Precision@K} \\
    \midrule 
    Random                          & 0.102 & 0.194 & 0.027 \\ 
    CLIP (Text)                     & 0.346 & 0.530 & 0.164 \\ 
    CLIP (Image)                    & 0.529 & 0.666 & 0.325 \\ %
    BLIP (Text)                     & 0.328 & 0.457 & 0.190 \\ 
    BLIP (Image)                    & 0.316 & 0.468 & 0.178 \\ 
    FLAVA (Text)                    & 0.276 & 0.446 & 0.137 \\ 
    FLAVA (Image)                   & 0.393 & 0.562 & 0.230 \\ 
    \textbf{\embedding{} (Image-Text)}  & \textbf{0.667} & \textbf{0.743} & \textbf{0.515} \\ %
    \bottomrule
  \end{tabular}
}
  \vspace{1mm}
  \caption{Search Retrieval Experiment. Evaluation is done on the top-10 retrieved product candidates. Graded Relevance considers the three-classed relevance label \{IR, MR, HR\} while Recall@K and Precision@K treats MR+HR as the binary positive label and IR as the negative label.}
  \label{table:retrieval}
  \vspace{-7mm}
\end{table}

\begin{table*}
  \begin{tabular}{clcc}
    \toprule
    \textbf{Evaluation Dataset} & \textbf{Model Variant} & \textbf{ROC-AUC} & \textbf{Norm LogLoss} \\
    \midrule
    \multirow{4}{*}{$NW$} & DCN baseline & .7731 $\pm$ .00006 & .8788 $\pm$ .00008\\
                        & DCN + product + query + similarity & .7741 $\pm$ .0001 & .8781 $\pm$ .0001 \\
                        &\textbf{DCN + product + query + similarity + purchase history} & \textbf{.7787 $\pm$ .0001} & \textbf{.8747 $\pm$ .0001} \\ 
                        & DCN + product + query + similarity + purchase history (random initialization) & .7731 $\pm$ .0002 & .8788 $\pm$ .0002 \\ 
    \hline
    \multirow{2}{*}{\shortstack{$NW$ $\cap$ $U_{PurcHist}$}} & DCN baseline & .7807 $\pm$ .0002 & .8706 $\pm$ .0002 \\
                                & DCN + product + query + similarity + purchase history & .7876 $\pm$ .0001 & .8653 $\pm$ .0001 \\
    \hline
    \multirow{2}{*}{\shortstack{$NW$ $\setminus$ $U_{PurcHist}$}} & DCN baseline & .7456 $\pm$ .0004 & .9183 $\pm$ .0006\\
                                & DCN + product + query + similarity + purchase history & .7475 $\pm$ .0006 & .9077 $\pm$ .0004 \\                          
    \bottomrule
  \end{tabular}
  \vspace{1mm} 
  \caption{Search Ranking Experiment. We collect the evaluation data from one week after the training data (called $NW$). Several evaluations were done to highlight different perspectives of the performance gain. The best candidate with product + query + purchase history embeddings is bold and its improvements over the baseline were statistically significant ($p<0.05$). Also, the candidate performed better for users with purchase history ($U_{PurcHist}$) by capturing a semantic profile of the users.}
  \label{table:ctr}
  \vspace{-4mm}
\end{table*}

\subsection{Ranking: Offline Experiments}

In the ranking task, we first present the problem formulation and the current model setup for the CTR prediction task on the search surface. Suppose that there is a potential product candidate eligible as ads when a user searches with a query string. The ranking model will take their (sparse, dense, cross) features as input and return a value between $[0, 1]$ which is interpreted as the probability of the user clicking the ads (impression of "sponsored product"). The model of choice in the current production environment is a Deep \& Cross Network variant \cite{wang2017deep, wang2021dcn} trained on binary labels of historical user click engagements. 

\begin{figure}[h]
  \centering
  \includegraphics[width=\linewidth]{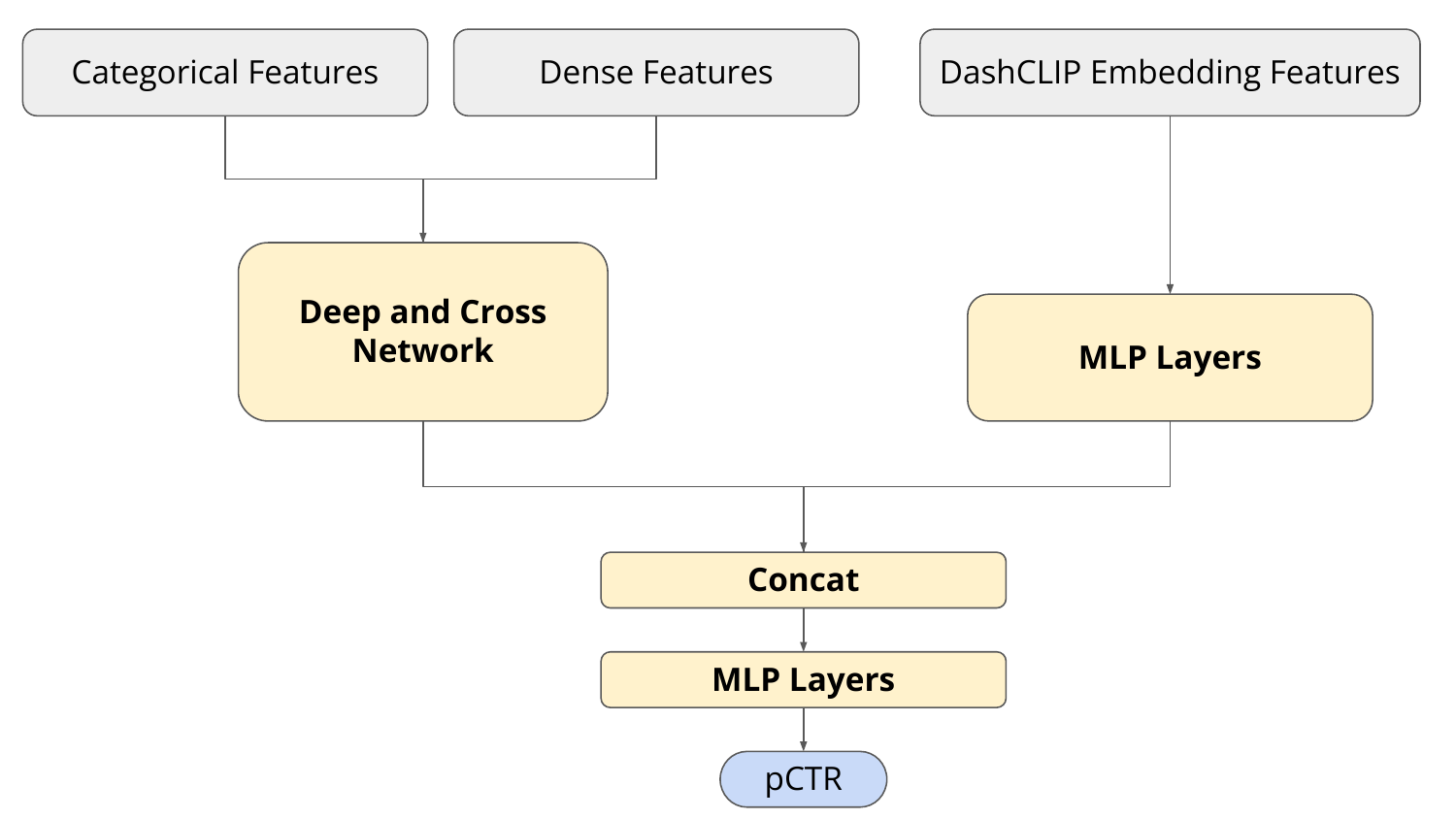}
  \caption{Model architecture for integrating the embedding features with the existing pCTR model. The outputs of the two-towers are concatenated and then passed through fully-connected layers to obtain the final click probability.}
  \label{fig:pctr_integration}
  \vspace{-4mm}
\end{figure}

Next, we illustrate the integration of the projected embeddings of the product $embed_p$, query $embed_q$, and their derived features with the ranking model. Our model architecture is shown in Figure \ref{fig:pctr_integration}: a DCN tower that behaves the same as the current production model, and a new tower that takes the product and query embeddings from \embedding{} and their derived features as input. The intermediate outputs from the two towers are concatenated and passed through a few fully connected layers before the final sigmoid layer that produces the output prediction in the range from 0 to 1. In the offline experiment stage, we iterated over different architecture designs and found that this approach, which separates the existing dense and sparse features from the new embedding-related features using the two-towers, achieved the best performance. This architecture promotes the crossing between the different embeddings before interacting them with the existing features. The derived features are of two types: the similarity between embeddings and representations of user engagement history on the platform. The former is the cosine similarity of the embeddings such as $cosine(embed_p, embed_q)$ and aims to capture the present user intent. One particular instance of the latter is the $consumer\_p84d\_purchased\_product$ which is a list of product ids that the user purchased in the last 84 days on the platform. Then it retrieves the product embeddings using the product ids as indices from the pre-computed embedding table. A mean pooling averages the retrieved embeddings, and then pooled vector is fed to the model. Similarly, $cosine(embed_p, pool(purchased\_product))$ captures how relevant the product is to the user's purchase profile.

\subsubsection{Search Ranking}
Table \ref{table:ctr} shows the offline performance of different models we trained with different architectures and features averaging over 5 runs (mean $\pm$ 1 std are reported). Models are trained with 7 months of users' click engagements and evaluated on the following week of data. Therefore, the train / evaluation dataset is split in the time dimension unless specified. From the first section of the table (Evaluation Dataset = $NW$), the new proposed model with the product, query embeddings, and derived purchase history (AUC = 0.7787) greatly improved over the DCN-only baseline (AUC = 0.7731) in terms of offline ROC-AUC metric. With the exact same architecture and feature set, we also evaluated the performance when the product and query embeddings were randomly initialized and gradient optimized along with the ranking model. The result shows that the model does not improve compared to the baseline indicating the necessity of \embedding{}'s training framework.

\subsubsection{Semantic User Profile}
The dimensional analysis is to quantify the gain of \embedding{} in terms of encoding the relevance between product and query versus representing the user profile. To break down overall improvement, product and query embeddings (AUC = 0.7741) partially drive the gain, while the purchase history further boosts the metric by a lot (AUC = 0.7787). Intuitively, while embeddings captures the semantic relatedness between the candidate and the search intent, the engagement history represents the user profile to provide further personalized recommendation. Therefore in Table \ref{table:ctr}, when we evaluate the best model candidate, it is observed that users with purchase history ($NW \cap U_{PurcHist}$) can benefit more than the users without any purchases ($NW \setminus U_{PurcHist}$), which confirmed the applicability of the embeddings to represent user interest.

\subsubsection{Feature Importance}
To understand the contribution of different features, we perform a feature importance analysis on the best model candidate using Captum \cite{kokhlikyan2020captum}, which is a model interpretability tool for PyTorch. There are implementations of many gradient and permutation-based algorithms to attribute model performance to features. We found out that the attribution is quite consistent across a few selected ones, including Integrated Gradients, Feature Ablation, and Feature Permutation. So, only the result of Feature Permutation is reported here. Among all features, query embedding, product embedding, and the user's purchase history of products in the past 84 days are the top-3 features, respectively, beating all existing dense and sparse features by a large margin. In particular, query embedding pushes a few query-based counting features further down in the importance list. So, embedding captures the relevance and represents the user engagement statistics of the search term.

\subsection{Ranking: Online Deployment}
\subsubsection{A/B Testing \& Rollout}
To evaluate the best model candidate in production, we created an online A/B experiment to serve real users on our platform. This consumer-bucketed test aimed to measure the business metric improvement of the new treatment model over the control DCN baseline. It was held in August 2024 for 10 days on the in-store surface. The experiment analysis showed that the new model increased the engagement rates for most of the top queries and categories, driving more revenue for the Sponsored Products advertisement and improving the relevance measure. The quantitative gain is reported in Table \ref{table:abmetric}. With the positive result of top-line metrics, the model is fully deployed to serve 100\% of the traffic by the end of the same month. Compared to labeling and training cost reported in Section \ref{cost}, the business revenue gain justified the large return on investment and deployment.

\begin{table}[htb!]
  \begin{tabular}{ccc}
    \toprule
    Click-through Rate & Conversion Rate & Revenue\\
    \midrule
    +3.73\% & +4.06\% & +1.46\% \\
    \bottomrule
  \end{tabular}
  \vspace{1mm}
  \caption{Top-line business metrics from A/B Experiment in August 2024. All reported values are statistically significant.}
  \label{table:abmetric}
  \vspace{-7mm}
\end{table}

\subsubsection{Online Serving} In order to retrieve the pre-trained embedding for online inference, we compared two viable strategies: feature fetching and model fusion.  For the former, we fetch the embedding features by doing product / user / query id lookups from the online feature store and feed the embeddings to the model. For the latter, we implant embedding tables directly in the model definition by extending the \textit{EmbeddingBagCollection} module of the \textit{torchrec} library to fuse the tables with the model. These strategies present a trade-off between network overhead induced by fat feature fetches of the actual embedding values vs memory constraint of the larger fused model size. Eventually we serve with the latter approach because latency will increase substantially due to fetches when the data and model are not co-located closely. For instance, the model fusing approach increases the p99 latency by less than 4ms while feature fetching typically consumes 5x more latency budget than model inference.

\section{E-commerce Applications}
\label{prelim-eval}


Given our goal to develop generalized embeddings for different applications, we explore \embedding{} in tasks beyond search recommendation. We pick aisle category prediction and product-query relevance prediction as two additional tasks. These tasks are important to improve the overall user experience on our platform and analogous tasks are prevalent across other e-commerce platforms: Accurate product categorization ensures seamless product discovery, and query-based relevance filtering is crucial to improve the user search experience. Further, the aisle-category prediction checks if the \embedding{} embeddings are aligned to our domain and product-query relevance tests the alignment between product and query encoders. Together, they perform a comprehensive evaluation of our model. We perform both qualitative and quantitative evaluations using the set-aside 2\% of the Query-Product Relevance Dataset with stratification by aisle category. The details of the tasks are as follows:


\noindent \textbf{Aisle Category}: We check if the embeddings are able to capture the aisle category of the products despite not being provided explicitly during the training. For qualitative evaluation, we plot the product embeddings after performing the t-SNE \cite{van2008visualizing} dimensionality reduction and annotate each point (product) with its ground truth category. For quantitative evaluation, we model this as a $n$-class classification task where the model predicts the category class.

\noindent \textbf{Product-Query Relevance}: This task checks whether the product-query relevance is captured by the embeddings. We qualitatively evaluate the relevance by plotting the distributions of cosine similarities between the embeddings of product query pairs. We also model this as a 3-class classification task where our classifier takes the pair of <query, product> embeddings as features to predict the relevance label.



\subsection{Aisle Category Prediction}
\label{prelim-eval-category}
We further perform an 80-20 stratified split on the unique product ids within the set-aside data to create the training and testing datasets. The dataset sizes are 13500 and 3375, respectively. For the baseline, we use the BLIP-14M checkpoint and compare it with \embedding{}. We train a simple classifier on top of the image-text Product Encoder by passing the last embedding output of the encoder through a hidden layer, followed by dropout and a final linear layer. We keep the encoder frozen and only train the added layers. 

\begin{figure}[h]
  \centering
  \includegraphics[width=\linewidth]{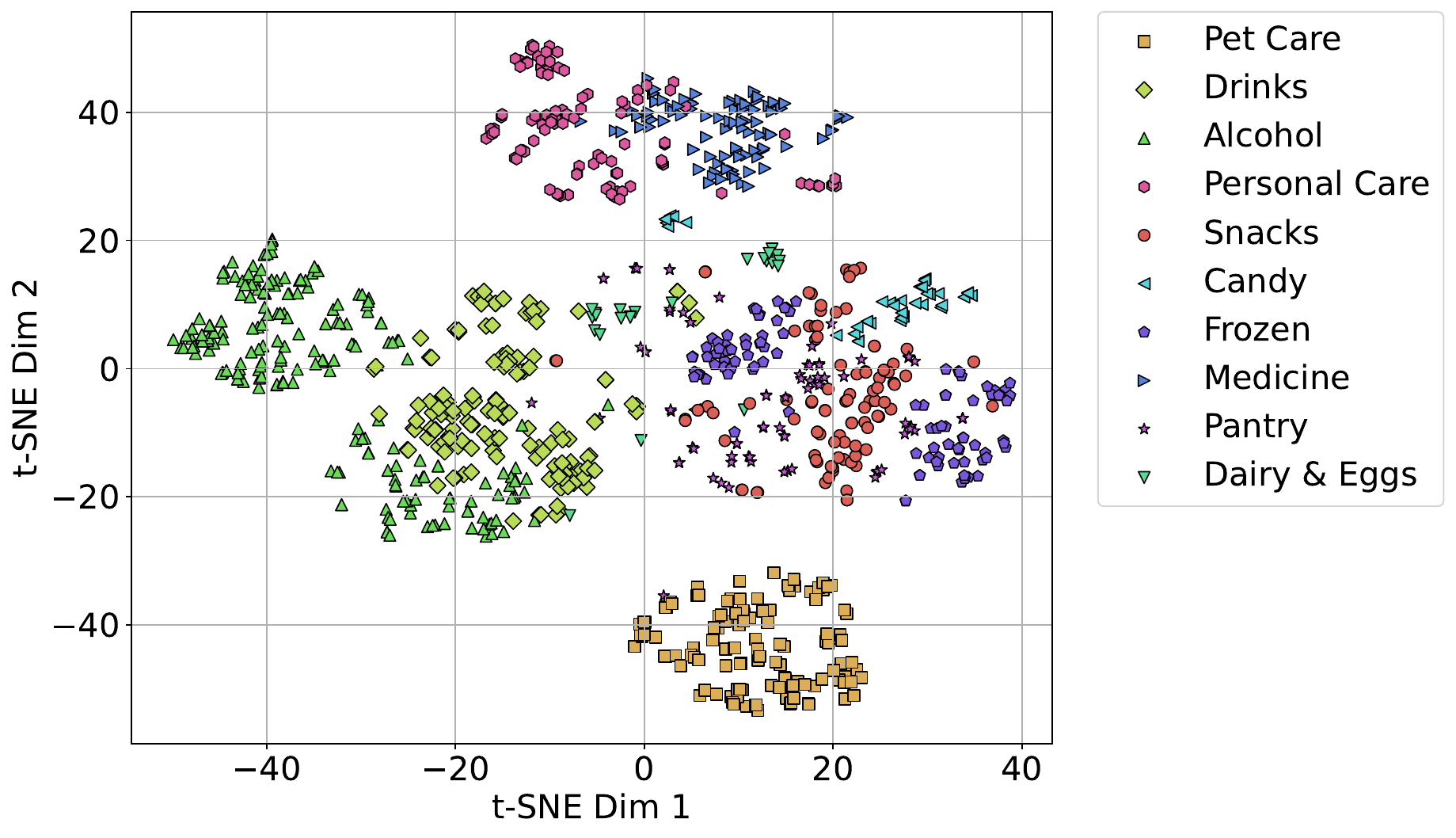}
  \caption{Scatter plot of product embeddings after t-SNE dimensionality reduction for top-10 aisle categories by frequency. Products from the same categories form clusters naturally. Similar clusters like Drinks and Alcohol are closer to each other. Cluster of unique categories like Pet Care is isolated from the majority mass.}
  \vspace{-1mm}
  \label{fig:tsne}
  \vspace{-4mm}
\end{figure}

We observe that our model performs significantly better with an average F1 score of 0.750 compared to 0.707 for the baseline. The baseline model tends to perform poorer on recall for low-support classes, whereas our model is able to limit this issue.
From the qualitative evaluation results shown in Figure \ref{fig:tsne}, we observe that the classes are naturally clustered, and semantically similar clusters are closer to each other.


\subsection{Product-Query Relevance Prediction}
\label{prelim-eval-relevance}
To prepare the dataset for this task, we first sample 10\% of data points from the set-aside dataset. Next, due to the class imbalance in the Query Product Relevance dataset from Section \ref{query-prod-dataset}, we randomly drop 50\% of IR class samples to remove easy negatives. Finally, we perform an 80-20 split on the obtained data to generate the train and test sets. The final sizes are 32577 and 8175, respectively. We again use the BLIP-14M checkpoint as the baseline. To build the classifier, we extract the multi-modal product embedding from the image-text encoder and concatenate the query embedding to pass it through a dropout layer followed by a classification layer. For the BLIP-14M, we encode the queries using the text encoder, and for \embedding{}, the query encoder is utilized to generate the query embedding. Again, we only train the added layers. 

\begin{figure}[h]
  \vspace{-2em}
  \centering
  \includegraphics[width=\linewidth]{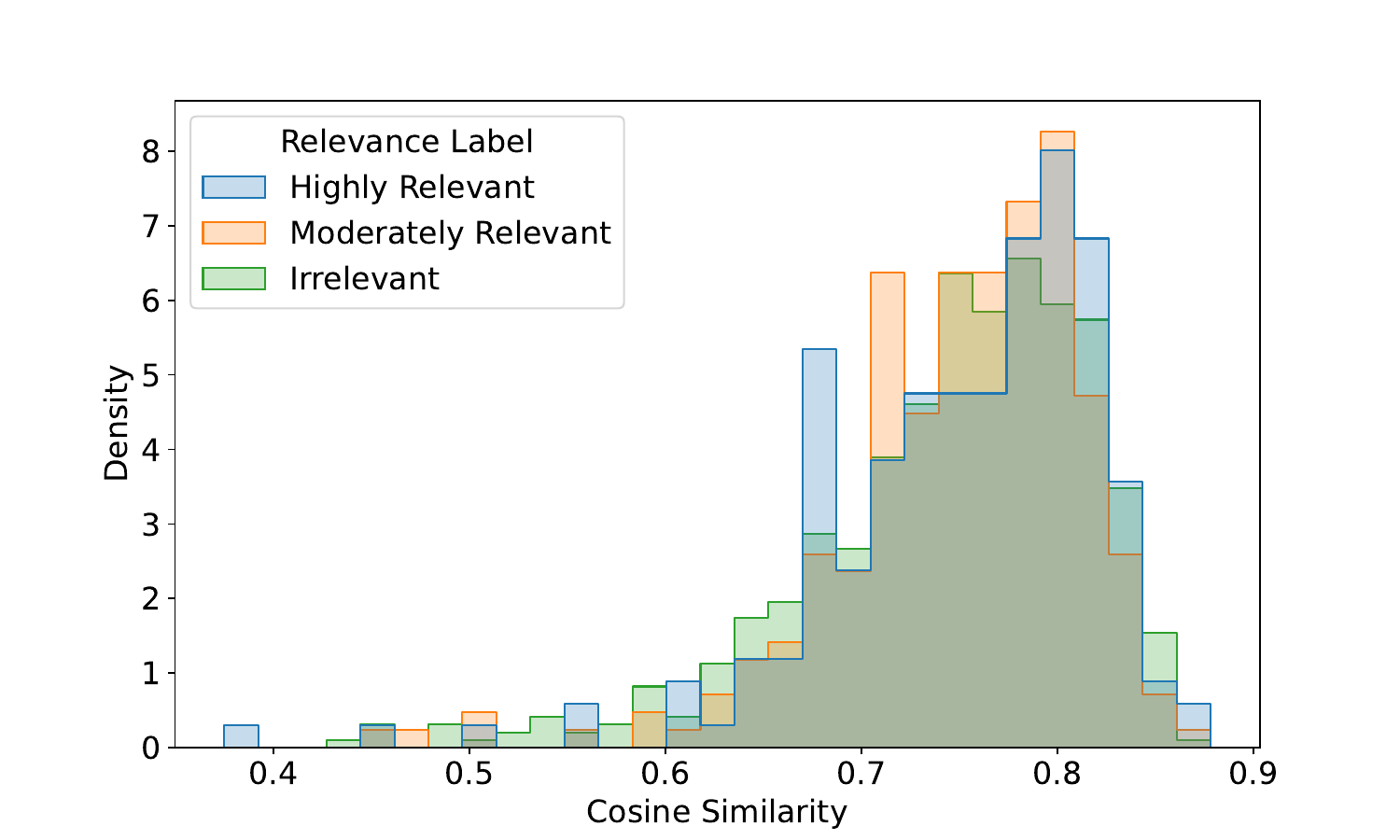}
  \includegraphics[width=\linewidth]{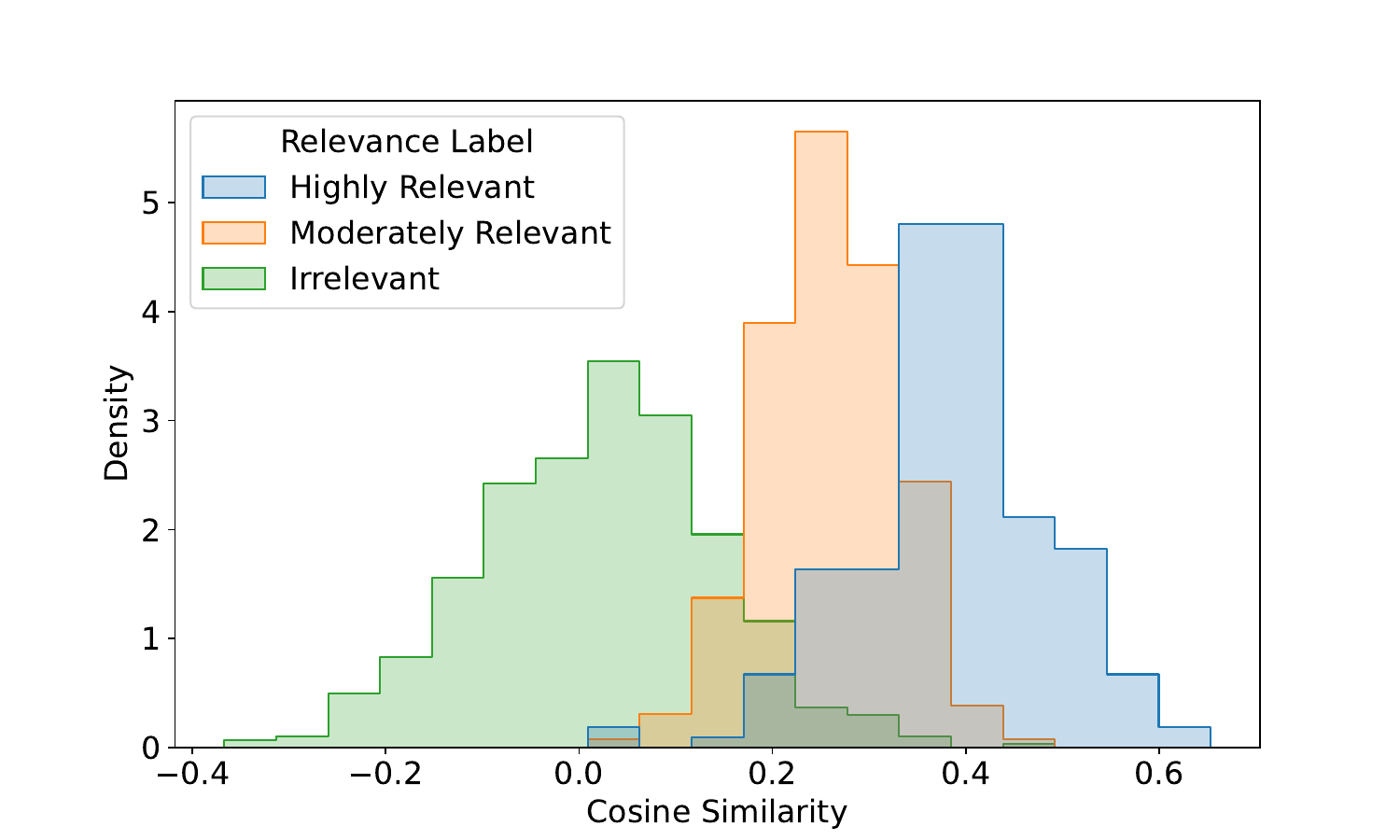}
  \caption{Distribution of cosine similarity between product and query embedding from off the shelf BLIP-14M (top) and \embedding{} (bottom). Our embedding is able to achieve a clear separation between the three relevance classes showing the effectiveness of PQC loss.}
  \label{fig:class_seperation_histogram}
  \vspace{-1mm}
\end{figure}

Our model achieves an average F1 score of 0.671 compared to 0.476 for the baseline. Similar to the first task, compared to our model, the baseline model achieves an extremely low recall (0.059) for the HR class due to its low frequency. From the qualitative evaluation shown in Figure \ref{fig:class_seperation_histogram}, \embedding{} is able to create significantly more separation between the three classes compared to the baseline. \\

\section{Conclusion}

Our proposed approach for generating multi-modal embeddings for products and search queries has broad applicability across e-commerce and social media search use cases. As foundational large language models and multi-modal models continue to evolve, our findings highlight that off-the-shelf models alone may not deliver optimal performance. We suggest that a representation of the entities should be built by pre-training on semantic data before the application-specific optimization.


The versatile nature of our \embedding{} framework has a lot of capabilities for different tasks with specific needs on the data modalities. For simpler tasks (such as category prediction), embeddings can be used directly out of the box. For more complex tasks, embeddings can serve as auxiliary and supporting information (such as search ranking). While our current focus has been on ranking tasks, future work will explore incorporating these embeddings into search retrieval, further enhancing the search experience by ensuring better query-product alignment. This adaptability underscores the broader potential of our approach in optimizing multi-modal representations for the e-commerce use-cases.

\section{Related Work}

\subsection{Pre-trained Embeddings: }
Contextualized representations, or embeddings, derived from pre-trained models have demonstrated state-of-the-art performance across various natural language \cite{devlin2018bert} and vision tasks \cite{dosovitskiy2020image}. 
Early works such as Word2Vec \cite{mikolov2013efficient} and Glove \cite{pennington2014glove} focused on learning word representations based on neighboring context and co-occurrence.
ELMo \cite{sarzynska2021detecting} improved upon these with an LSTM-based architecture which enabled context-dependent word representations.

The transformer architecture \cite{vaswani2017attention} gave rise to popular pre-trained language models like BERT \cite{devlin2018bert} and RoBERTa \cite{liu2019roberta}. Subsequent works leveraged this architecture to learn sentence-level embeddings \cite{reimers2019sentence} \cite{gao2021simcse}. More recently, approaches have expanded to learn vision \cite{radford2021learning, dosovitskiy2020image} and multi-modal \cite{li2021align, li2022blip, li2023blip} representations, enabling zero-shot transfer learning for vision-language tasks.

Use of general datasets while training pre-trained models often leads to subpar performance on specific use cases, consequently requiring domain adaptation. In e-commerce and social media applications embeddings are widely used for personalized ranking and search. \citet{nigam2019semantic} learned semantic user representations using activity logs. MARN \cite{li2020adversarial} uses an adversarial network to generate multi-modal representations and achieved SoTA performance on several CTR prediction tasks. \citet{baltescu2022itemsage} utilizes a combination of different engagement signals in a multi-task learning setup. Beyond these, numerous studies \cite{xv2022visual, zhang2024scaling, pancha2022pinnerformer, zhang2020general} have tried to learn product and user representations on online platforms.


\subsection{CTR Prediction: } CTR prediction is crucial for e-commerce platforms and aims to predict the probability of a user clicking an impression. Earlier works relied on traditional machine learning methods including logistic regression \cite{richardson2007predicting, kumar2015predicting}, kernel SVM \cite{chang2010training}, Factorization Machines (FM) \cite{rendle2010factorization, juan2016field, pan2018field}, and gradient boosting \cite{ling2017model} to predict CTR using different features. Deep Learning models lead to improved performance due to their ability to learn complex non-linear interactions: 
Product-based Neural Networks (PNN) \cite{qu2016product} captured category interactions though embedding layers. The Wide and Deep architecture \cite{cheng2016wide} used a combination of wide cross-product transformations for feature interactions memorization and deep neural network for generalization, training both components jointly. Deep and Cross Network (DCM) \cite{wang2017deep} reduced the need for manual feature engineering by introducing the cross-network. DeepFM \cite{guo2017deepfm} combined DNNs with FMs to improve feature interaction. FiBiNET \cite{huang2019fibinet} used a bilinear function to dynamically learn feature interactions and demonstrates SoTA performance combining with both shallow and deep models. Finally, DCNV2 \cite{wang2021dcn} improved on DCN, allowing large-scale learning in industrial settings. While many of these models emphasize architectural enhancements, our approach remains architecture-agnostic, focusing on generating semantic embeddings that can be seamlessly integrated into any architecture.

\bibliographystyle{ACM-Reference-Format}
\bibliography{references}

\appendix



\end{document}